\documentclass[a4paper,11pt]{article}
\usepackage{pos}
\usepackage{amsmath}
\usepackage{graphicx}

\newcommand{\asymrel}[4]{%
  \smash{\left(
    \vcenter{\offinterlineskip
      \halign{%
        \hfil$\scriptscriptstyle##$\hfil&
        $\scriptscriptstyle.##$\hfil\cr
        +#1&#2\%\cr
        \noalign{\kern1.0pt}
        -#3&#4\%\cr
      }%
    }
  \right)}%
}

\title{Measurements of lepton-flavour universality in semileptonic $B$ decays at Belle and Belle~II}

\author*[1]{Andreas Warburton,}
\onbehalf{on behalf of the Belle and Belle~II Collaborations}

\affiliation{Rutherford Physics Building, McGill University\\
   Montr\'eal, Canada}

\emailAdd{andreas.warburton@mcgill.ca}

\note{Supported by the Natural Sciences and Engineering Research Council of Canada}

\renewcommand{\printHeadAuthors}{A.~Warburton}

\abstract{Tests of lepton-flavour universality in semileptonic $B$-meson decays provide sensitive probes of physics beyond the Standard Model. 
The Belle and Belle~II experiments, respectively operating on the KEKB and SuperKEKB colliders in Japan, have collected $e^+e^-\to B\bar{B}$ collisions at a center-of-mass energy corresponding to the $\Upsilon(4S)$ resonance in samples with a combined integrated luminosity exceeding 1~ab$^{-1}$.
These data, with low particle multiplicity, constrained initial-state kinematics, and excellent lepton identification, provide an ideal environment to study lepton-flavour universality in semileptonic decays of $B$ mesons.  New results on the ratios of semitauonic $B$-decay rates compared to analogous light-lepton rates are presented.  These include new Belle~II measurements of branching-fraction ratios $R(D^{(*)})$ involving exclusive $B\to D^{(*)}\tau\nu$ and $B\to D^{(*)}\ell\nu$ decays
using semileptonic or hadronic reconstruction of the accompanying $B$-meson tag.
Belle~II and Belle measurements of angular observables in $B\to D^*\ell\nu$ modes that test light-lepton universality and the Lorentz structure of the decay are also presented. The measurements are statistically limited and individually consistent with Standard Model expectations. The new $R(D^{(*)})$ results are also compatible with experimental world averages, which continue to exhibit tension with Standard Model predictions. Belle~II has now accumulated an $\Upsilon(4S)$ data sample exceeding that of Belle, offering substantial opportunities for new analysis approaches and heightened precision.}

\FullConference{XVI International Conference on Beauty, Charm, Hyperons in Hadronic Interactions (BEACH 2026)\\
7--12 June 2026\\
Firenze, Italy\\}

\begin{document}
\maketitle

\section{Lepton-flavour universality in semileptonic $B$ decays}

In the Standard Model (SM), the charged-current couplings of the three flavours of charged leptons to the $W$ boson are universal, apart from effects associated with their different masses. Semileptonic $B$-meson decays therefore provide precise tests of lepton-flavour universality (LFU)~\cite{CDFWtau,ATLASWtauMu,CMSWbranching}, with sensitivity to possible additional charged-current interactions. Of particular interest are the branching-fraction ratios
\begin{equation}
 R(D^{(*)}) \equiv
 \frac{\mathcal{B}(\bar B\to D^{(*)}\tau^-\bar\nu_\tau)}
      {\mathcal{B}(\bar B\to D^{(*)}\ell^-\bar\nu_\ell)},\qquad \ell=e,\mu,
\end{equation}
for which many experimental and theoretical uncertainties cancel. The semitauonic and light-lepton modes share the same underlying $b\to c$ quark transition, while the larger $\tau$ mass enables sensitivity to non-SM scalar, vector, and tensor interactions~\cite{LondonMatias,CrivellinHoferichter}. The combined experimental measurements of $R(D)$ and $R(D^*)$ show tension with SM predictions~\cite{HFLAV}, motivating both improved rate measurements and complementary studies of differential and angular observables.

The angular distributions of $B\to D^*\ell\nu$ decays provide information beyond that contained in total rates. The decay can be described in terms of the recoil variable $w=(m_B^2+m_{D^*}^2-q^2)/(2m_Bm_{D^*})$ and three helicity angles. Distributions in these variables are sensitive to the Lorentz structure of the underlying interaction. Comparing angular observables between electron and muon modes gives direct tests of light-lepton universality, with substantial cancellation of common theoretical and experimental uncertainties~\cite{Bobeth:2021}.

\section{Belle~II and $B$ tagging}

Located in Tsukuba, Japan, Belle~II~\cite{Belle2TDR} is operating at the SuperKEKB asymmetric-energy $e^+e^-$ collider~\cite{SuperKEKB}, primarily at the $\Upsilon(4S)$ resonance. The well-defined initial state and the clean production of $B\bar B$ pairs are particularly advantageous for final states containing neutrinos. One tagging $B$ meson can be reconstructed to constrain the kinematics of the other $B$ meson on the signal side. The resulting tagging strategies span a range from inclusive methods, with high efficiency but relatively limited knowledge of the tag-side kinematics, through semileptonic tagging, to exclusive hadronic tagging, which provides higher purity at a cost of efficiency.

The following results illustrate the complementarity of these approaches. Two Belle~II measurements determine $R(D^{(*)})$ using semileptonic and hadronic tags, while angular measurements from Belle~II and its predecessor Belle use hadronic tagging to exploit the constrained event kinematics.

\section{Measurements of $R(D^{(*)})$ at Belle~II}

\subsection{Semileptonic-tag measurement}

\begin{figure}[t]
\centering
\includegraphics[width=0.48\textwidth]{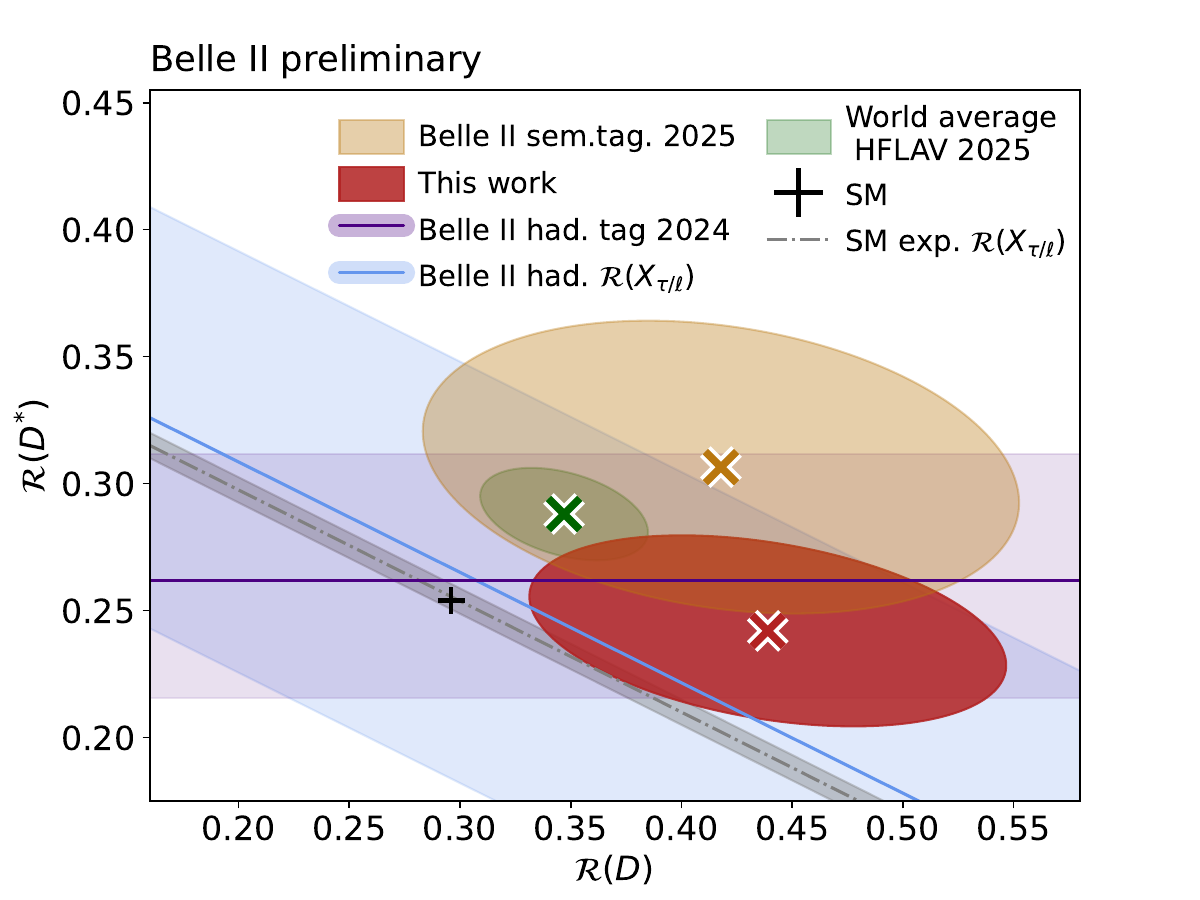}
\caption{The new Belle~II hadronic-tag (red) $R(D)$ and $R(D^*)$ results~\cite{Belle2Hadtag} in the $R(D)$--$R(D^*)$ plane, with comparisons to the experimental world average (green)~\cite{HFLAV} and SM expectation (black marker), as well as the recent semileptonic-tag results (yellow)~\cite{Belle2SLtag} and earlier Belle~II measurements (bands)~\cite{Belle2OldRDstar,Belle2RXtau2024}.}
\label{fig:had}
\end{figure}

A multivariate classifier separates three event categories: semitauonic $B\to D^{(*)}\tau\nu$ signal, light-lepton $B\to D^{(*)}\ell\nu$ normalization decays, and background. The classifier uses five input variables, with strong discrimination supplied by the kinematic consistency of the semileptonic tag. Signal and normalization yields are determined in a two-dimensional binned template fit to combinations of the classifier outputs. Four channels, $D^+e^-$, $D^+\mu^-$, $D^{*+}e^-$, and $D^{*+}\mu^-$, are fitted simultaneously.
The measured ratios, plotted in Fig.~\ref{fig:had}, are
\begin{align}
 R(D^+) &= 0.418 \pm 0.075\,(\mathrm{stat}) \pm 0.056\,(\mathrm{syst}),\\
 R(D^{*+}) &= 0.306 \pm 0.035\,(\mathrm{stat}) \pm 0.018\,(\mathrm{syst}).
\end{align}
The dominant systematic-uncertainty contribution to $R(D^+)$ is the finite simulated-sample size ($\pm 7.9\%$), with additional key contributions from gap-mode\footnote[2]{Gap modes, of the form
$\bar{B}\to D^{(*)}\eta\ell^-\bar{\nu}_\ell$, are simulated to account for the difference between the total inclusive semileptonic decay rate and the sum of observed exclusive rates.}
 modelling $\asymrel{3}{4}{9}{1}$, lepton-identification $\asymrel{6}{1}{3}{9}$, and fake-rate $\asymrel{3}{7}{2}{0}$ corrections. Accounting for the correlation between the two ratios, the result differs from the SM expectation by $1.7$ standard deviations ($\sigma$) and is compatible with experimental world-average values~\cite{Belle2SLtag}.

\subsection{Hadronic-tag measurement}

The principal background challenge is from $B\to D^{**}\ell\nu$ and related semileptonic modes, which can enter the signal region and are not all precisely reconstructed. Events are required to contain no additional charged tracks after reconstruction of the tag and signal candidates, and signal and background modelling are validated in dedicated sideband regions. The signal yield is extracted from a two-dimensional binned extended maximum-likelihood fit to the squared missing mass $M^2_{\rm miss}\equiv (p_{e^+e^-}-p_{B_{\rm tag}}-p_{D^{(*)}}-p_\ell)^2$ and the residual energy $E_{\rm ECL}$ in the electromagnetic calorimeter. These variables exploit the additional semitauonic neutrinos and the completeness characteristic of correctly reconstructed signal events.
The measured ratios, depicted in Fig.~\ref{fig:had}, are
\begin{align}
 R(D) &= 0.439 \pm 0.055\,(\mathrm{stat}) \pm 0.046\,(\mathrm{syst}),\\
 R(D^*) &= 0.242 \pm 0.019\,(\mathrm{stat}) \pm 0.016\,(\mathrm{syst}).
\end{align}
This constitutes the most precise determination of $R(D^{(*)})$ with hadronic tagging to date~\cite{Belle2Hadtag}, superseding Ref.~\cite{Belle2OldRDstar}. 
The leading systematic-uncertainty contributions to $R(D^*)$ [$R(D)$] arise from the simulation sample size, $\pm 4.8\%$ [$\pm 8.4\%$], and the gap-mode branching fractions, $\pm 2.7\%$ [$\pm 3.1\%$].
The combined $R(D)$--$R(D^*)$ measurement differs from the SM by $1.5\sigma$ and from the experimental world average by $1.3\sigma$~\cite{Belle2Hadtag}.

\vspace*{-0.05in}

\section{Angular tests of lepton-flavour universality}

\subsection{Angular asymmetries at Belle~II}

\begin{figure}[t]
\centering
\includegraphics[width=0.82\textwidth]{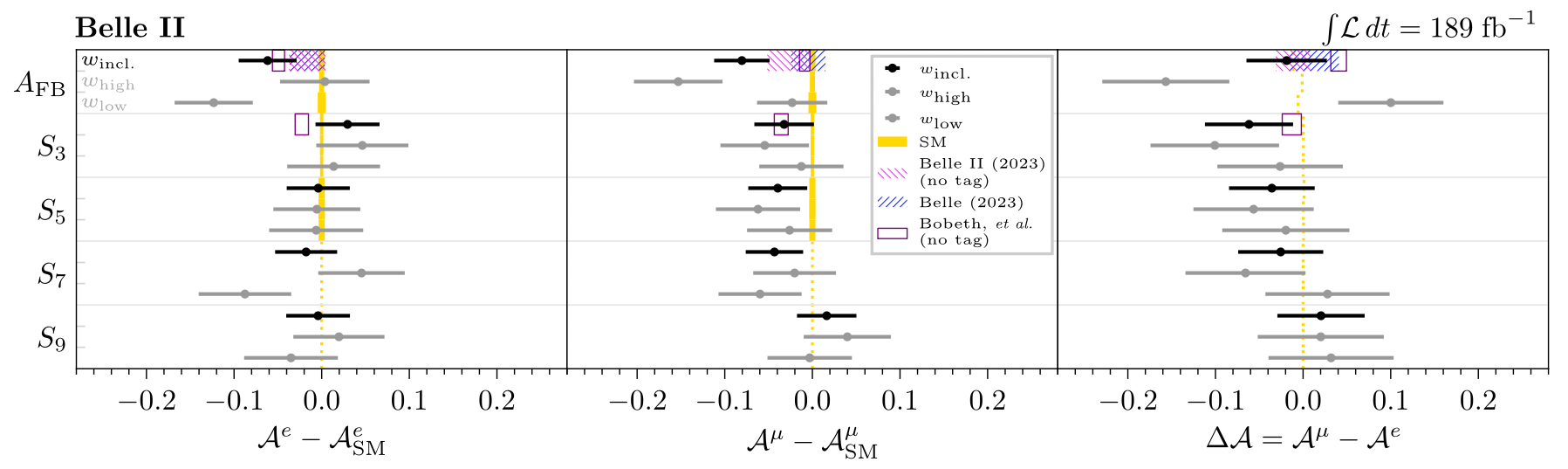}
\caption{Belle~II angular asymmetries and electron--muon differences in $B^0\to D^{*-}\ell^+\nu_\ell$ decays, shown for inclusive and recoil-binned samples~\cite{Belle2Angular}. The asymmetry observables $A_{\rm FB}$ and $S_i$ are proportional to normalized angular coefficients $\hat{J}_i$ drawn from the full twelve-term differential decay rate.}
\label{fig:asym}
\end{figure}

Using a 189~fb$^{-1}$ sample, signal yields of the modes $B^0\to D^{*-}e\nu$ and $B^0\to D^{*-}\mu\nu$ are extracted separately in two angular regions and in recoil intervals, including low- and high-$w$ regions as well as the inclusive sample. The largest systematic uncertainty is due to the finite simulated samples used for efficiency and bin-migration corrections. Joint tests of the LFU-sensitive differences $\Delta A_{\rm FB}$, $\Delta S_3$, and $\Delta S_5$ give $p=0.57$ for the inclusive recoil range and $p=0.13$ for the two recoil subranges. The results, illustrated in Fig.~\ref{fig:asym}, provide no evidence for LFU violation~\cite{Belle2Angular}.

\subsection{Angular coefficients at Belle}

\begin{figure}[t]
\centering
\includegraphics[width=0.78\textwidth]{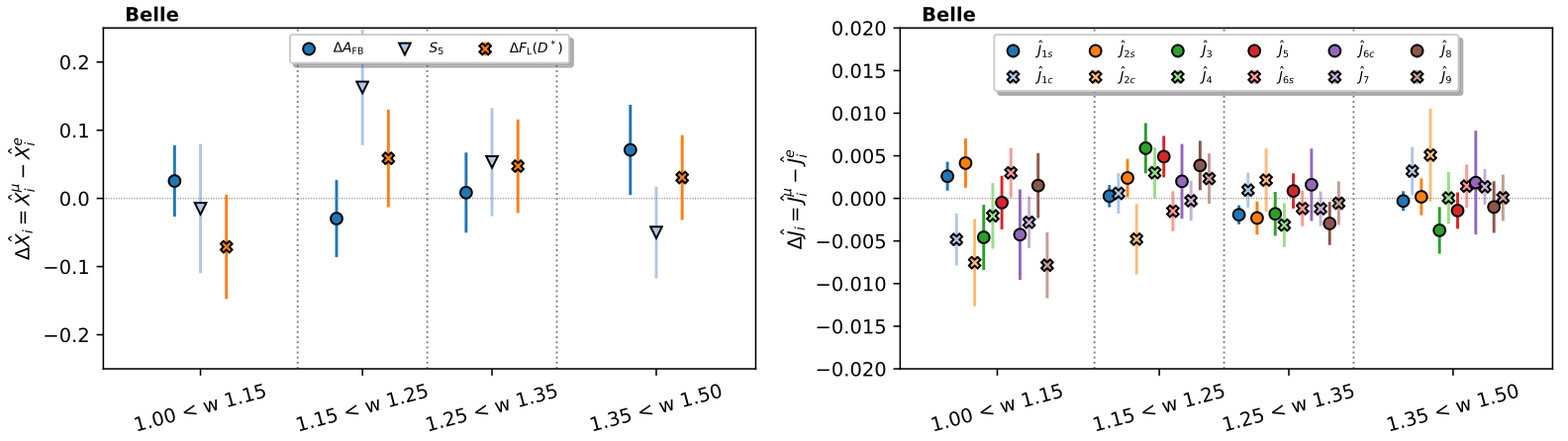}
\caption{LFU-sensitive observables (left) derived from angular coefficients (right) measured by Belle in $\bar B\to D^*\ell\bar\nu_\ell$ decays, expressed as differences between their determinations using
muon and electron final states. No significant deviation from the SM expectation is observed~\cite{BelleAngular}.}
\label{fig:coeff}
\end{figure}

In a study of all four decay modes of $B\to D^*\ell\nu$, with $B^{\pm,0}$ and $\ell = e,\mu$, the Belle experiment used its final 711~fb$^{-1}$ dataset to
measure the twelve angular coefficients $\hat{J}_i$ describing the fully differential decay distribution by simultaneously fitting 576 phase-space regions based on 36 bins in the three angular variables, four bins of recoil $w$, and the four decay modes. Differences between the $e$ and $\mu$ observables test LFU; for the complete set of angular-coefficient differences, the comparison with the SM gives $\chi^2/\mathrm{ndf}=41/48$ ($p=0.76$). The measured LFU-sensitive angular-coefficient combinations are shown in Fig.~\ref{fig:coeff}; no deviation from SM expectations is observed~\cite{BelleAngular}.

\vspace*{-0.18in}

\section{Summary and outlook}

Belle and Belle~II provide a broad program of LFU tests in semileptonic $B$ decays. Building on past measurements~\cite{Belle2OldRDstar,BelleRD2015,BelleTauPol2017,BelleSL2020,BelleQ2Mom2021,Belle2RXemu2023,Belle2RXtau2024}, the recent Belle~II $R(D^{(*)})$ results use complementary semileptonic- and hadronic-tag strategies, while angular measurements at Belle~II and Belle test $e$--$\mu$ universality and provide sensitivity to the Lorentz structure of possible non-SM interactions.
None of the four measurements individually exhibits a significant deviation from SM expectations, although the combined world-average \(R(D^{(*)})\) results continue to motivate further investigation.

Belle~II is increasing its data sample and has surpassed the integrated $\Upsilon(4S)$ sample accumulated by Belle. Larger datasets, improved reconstruction and tagging methods, and additional decay modes will allow substantially more precise rate and angular measurements. The combination of increasingly precise $R(D^{(*)})$ values with differential observables will be important for determining whether the present tensions persist and, in the event of a deviation, for identifying the structure of the underlying physical interactions.

\end{document}